\documentclass[reprint,
amsmath,amssymb,
 aps,
prl,
raggedbottom,
]{revtex4-2}

\usepackage{graphicx}\usepackage{dcolumn}\usepackage{bm}\usepackage{hyperref}\usepackage{tabularx}
\usepackage{float}
\usepackage{placeins}

\begin{document}

\title{Quantum Monte Carlo study of positron lifetimes in solids}

\author{K. A. Simula}
\author{J. E. Muff}
\author{I. Makkonen}\affiliation{Department of Physics, P. O. Box 43, FI-00014 University of Helsinki, Finland
}

\author{N. D. Drummond}
\affiliation{
Department of Physics, Lancaster University, Lancaster LA1 4YB, United Kingdom 
}

\date{\today}

\begin{abstract}
We present an analysis of positron lifetimes in solids with unprecedented depth. Instead of modeling correlation effects with density functionals, we study positron-electron wave functions with long-range correlations included. This gives new insight in understanding positron annihilation in metals, insulators and semiconductors.
By using a new quantum Monte Carlo approach for computation of positron lifetimes, an improved accuracy compared to previous computations is obtained for a representative set of materials when compared with experiment. Thus we present a method without free parameters as a usefulalternative to the already existing methods for modeling positrons in solids. 
  
\end{abstract}

\pacs{Valid PACS appear here}\maketitle

Positron annihilation is an elementary component of quantum electrodynamics \cite{ferrell1956}. Measurable annihilation parameters such as positron annihilation rate and momentum density of annihilation radiation are governed by many-body interactions between the positron and the surrounding electronic system. Theoretical predictions and experimental measurements can provide a good match for few-electron systems, for example the binding energies of positrons bound to molecules can be reproduced \cite{hofierka2021}.
Theoretical studies of homogeneous electron-positron systems provide a starting point for understanding positron annihilation in more complex systems \cite{boronski1986,drummond2011}, but real, inhomogenous systems are nevertheless often problematic. Better description of the correlations is needed to improve theory and applicability of positron physics. 

Positron annihilation spectroscopy \cite{tuomisto2013} is a powerful, nondestructive method for studying systems including metals, alloys, semiconductors and insulators \cite{krause1999}, polymers and soft matter \cite{pethrick1997,Jean2003}, and porous materials \cite{gidley2000}. 
The method involves injecting positrons into a sample where they first thermalize rapidly, and then either continue diffusing as a delocalized wave or become trapped in void-like open volumes. Eventually each positron annihilates with an electron, emitting two detectable 511 keV $\gamma$ photons. As the positrons are very sensitive to open volumes within a sample, measured lifetime components and their intensities provide information on the size and concentration of vacancies trapping the positrons \cite{tuomisto2013}.

Positron lifetime techniques can be combined with, for example, \textit{in situ} irradiation \cite{Segercrantz2017} or optical illumination \cite{Maki2011} to study damage production and various defect properties. Pulsed slow positron beams \cite{Schodlbauer1987} can be used for lifetime studies of thin films or scanning surfaces with a micrometer-scale lateral resolution \cite{david2001}. Also the study of pore size distribution in thin films is possible \cite{gidley2000}.
On the other hand, the local electronic momentum density of the annihilation site is directly connected to the Doppler broadening of the annihilation $\gamma$ radiation, enabling the study of the chemical surroundings of lattice defects \cite{tuomisto2013} or Fermi surfaces of metals \cite{Kontrym-Sznajd2009}.

Theoretical models associate measured lifetime components to microscopic traps in the material. Single-particle models, such as Hartree-Fock (HF) theory, ignore many-body correlation effects between particles. Such models give smaller positron-electron overlaps, overestimating the positron lifetimes by nearly an order of magnitude \footnote{In the HF theory the particles are uncorrelated and the contact pair correlation function is $g(0)=1$. However, in correlated simulations the value increases as demonstrated in Fig. 1.}. Better description of positrons is provided by two-component density functional theory (DFT) \cite{boronski1986}, where the many-body correlation effects are included in approximate correlation functionals of the particle densities. In the local density approximation (LDA), the electron-positron correlation energies are obtained from functionals of the local densities, and the electron-positron overlap enhanced by correlations is computed with another local functional, the so-called enhancement factor.
Generalized gradient approximations (GGA) \cite{barbiellini1995,kuriplach2014,barbiellini2015} can improve lifetime calculations in comparison to the underlying LDA parametrizations.

Despite successes with LDA or GGA  functionals, DFT is an insufficient theory for positron annihilation studies. A correlation functional may accurately predict the lifetime in some systems while failing in others. The lifetimes by different functionals in a given system can differ  even by $30$ ps (see below). Moreover, there is no \textit{a priori} means to determine a functional best suited for a given task but experimental benchmarking is needed; and the construction of a functional can require both higher level calculations, such as quantum Monte Carlo calculations \cite{drummond2011}, and fits to experimental data \cite{barbiellini2015}. 

For a given positron lifetime setup, the statistical accuracy achievable for a defect-free sample is on the order of $1$~ps.
In general DFT does not reach this accuracy. Moreover, local or semilocal correlation functionals are bound to fail in solid-state systems with large open volumes or surfaces. Thus often it is not possible to adequately assign the components of the measured lifetime spectra to different microstructures of the sample. Development of a practical many-body theory that is accurate and able to describe complex correlation effects has been a long-standing problem in the field of positron physics and its applications. More accurate many-body theory could improve the applicability of positron annihilation and provide new research areas. For example, the correlation functionals could be evaluated and improved using a more descriptive theory. A many-body theory also enables the study of positrons in systems with multideterminantal nature, impossible with current methods but unavoidable with many lattice defects \cite{hood2003}.

Importantly, the dependence of positron lifetime on lattice vibrations is a theoretical question unaddressed to date, and should clearly be studied for complete understanding of positron annihilation in solids. 

We present quantum Monte Carlo (QMC) as a new method, devoid of free parameters, for simulation of positrons in solids. To our knowledge, this is the first QMC study of positrons in real crystals, although such studies exist for molecules \cite{kita2009,kita2010} and electron gases \cite{drummond2011}. 
Calculation of positron lifetimes in the perfect bulk of materials is the first and the most critical test for benchmarking how well the electron-positron correlations in inhomogeneous solid-state systems can be described before we move on to other experimentally relevant quantities such as the momentum density of annihilating pairs. Annihilation studies in defect systems, such as vacancies, in which also the detailed ionic structure in the presence of the trapped positron poses a challenge, can follow after we have validated the capability of QMC to describe correlations in defect-free lattices.

We perform a detailed study of finite size effects involved and how to best describe annihilation with core electrons. Besides QMC we study vibrational effects on positron lifetime. 

The studies were performed for C and Si in the diamond structure, body-centered cubic Li, and wurtzite AlN, a set that consists of materials of past and present interest in the field of positron annihilation spectroscopy and includes insulators (C), semiconductors (AlN, Si) and metals (Li). The choice of the test set was limited by the need of experimental reference data and by the available pseudopotentials. Si and AlN have well-known experimental reference results. C is a less correlated system and expected to be easy to model with QMC. The positron lifetime in Li has been overestimated by many state of-the-art two-component DFT correlation functionals, making the results by QMC theoretically interesting.

We use variational and diffusion Monte Carlo (VMC and DMC) methods \cite{ceperley1980,foulkes2001} as implemented in the \textsc{casino} code
\cite{needs2009,needs2020}. 
The fermion-sign problem is treated in DMC by imposing a fixed-node approximation \cite{anderson1975} that constrains the nodal surface of the wave function to be that of the VMC-optimized wave function.

The VMC many-body wave functions are represented as Slater-Jastrow (SJ) or Slater-Jastrow-backflow (SJB) wave functions \cite{jastrow1955,rios2006}. The former is a product of single-particle Slater determinants and a Jastrow factor. The determinants fix the nodal surface, and the Jastrow factor is a parametrized function describing the interparticle correlations. The SJB wave function goes beyond the single-particle SJ nodal surface by introducing parametrized shifts into the particle coordinates.
Optimizing backflow parameters both increases variational freedom in VMC and reduces the DMC fixed-node error.  

The trial wave functions can be written for a system with one positron as 
\begin{align}
\label{equation:slater-jastrow wf, qualitative}
\Psi_{SJ}(\mathbf{R}) &= e^{J(\mathbf{R})}\left[ \phi^l(\mathbf{r}_{i\uparrow}) \right]\left[ \phi^m(\mathbf{r}_{j\downarrow}) \right]\phi(\mathbf{r}_+),\\
\Psi_{SJB}(\mathbf{R}) &= e^{J(\mathbf{R})}\left[ \phi^l(\mathbf{r}_{i\uparrow} -\boldsymbol{\xi}_{i\uparrow}(\mathbf{R})) \right]\left[ \phi^m(\mathbf{r}_{j\downarrow}-\boldsymbol{\xi}_{j\downarrow}(\mathbf{R})) \right]\nonumber\\ 
&\times \phi(\mathbf{r}_+-\boldsymbol{\xi}_+(\mathbf{R})), \nonumber
\end{align}
where  $\mathbf{r}_\uparrow$, $\mathbf{r}_\downarrow$ and $\mathbf{r}_+$ denote the positions of up- and down-spin electrons and  the positron, respectively. $N$ is the number of particles in the system. $\mathbf{R}$ is a $3N$-dimensional vector of the particle coordinates. $J(\mathbf{R})$ and $\boldsymbol{\xi}(\mathbf{R})$ are the Jastrow exponent and backflow displacement, respectively, parametrized with respect to different spin groupings. The $\phi$-functions are single-particle Kohn-Sham orbitals \cite{kohn1965}, computed with DFT using Quantum \textsc{espresso} \cite{giannozzi2009} and our own positron package \cite{torsti2006}. We assume that the delocalized positron density does not affect the average electron density and take the zero-positron-density limit of the e-p correlation energy functional \cite{MakkonenPRB2006}. The $[...]$-signs denote Slater determinants over the orbitals. The Perdew-Burke-Ernzerhof \cite{perdew1996} GGA and Boro\'nski-Nieminen \cite{boronski1986} LDA functionals were used to solve electron and positron orbitals, respectively. The orbitals were in a localized B spline, or blip, basis \cite{alfe2004} (see Supplemental material). 

Periodic boundary conditions generalize the definition of orbitals in Eq.~(\ref{equation:slater-jastrow wf, qualitative}). Each orbital of band $j$ at wave vector $\mathbf{k}$ $\phi^j_\mathbf{k}(\mathbf{r})$ is of the form $u_\mathbf{k}^j(\mathbf{r})e^{i\mathbf{k}\cdot\mathbf{r}}$, i.e., a lattice periodic function $u$ multiplied by a plane-wave exponential, according to Bloch's theorem. We use twist averaging, i.e. average results computed in grid of Bloch $\mathbf{k}$-vectors, but the positron orbital is always chosen from the minimum of the parabolic positron band ($\mathbf{k}=0$), as we focus on a single thermalized positron in an infinite lattice.

The Jastrow factor contains terms representing 1, 2, and 3-body correlations \cite{drummond2004}, and the backflow function contains 1 and 2-body terms. The Jastrow factor is optimized with a variance minimization method (except when core electrons are included we use energy minimization, see below) \cite{drummond2005} and the backflow is optimized together with the Jastrow factor with an energy minimization algorithm \cite{umrigar2007}.

Multiple finite-size effects bias our simulations. The long-range correlations are not described correctly by finite cell sizes, and quasirandom finite-size noise arising from the forcing of Friedel oscillations to be commensurate with the simulation cell is difficult to remove from the calculation \cite{needs2020}. Momentum integrals are treated as discrete sums, increasing the kinetic energy bias \cite{holzmann2016}. Noise due to discrete momentum grid is reduced by twist averaging \cite{lin2001}.
The relaxation energies are computed by fitting computed energies to the twist vectors~ \cite{needs2020} (see Supplemental material).
Finite-size effects due to long-range interactions can be reduced by increasing the simulation cell size.  Coulombic interactions are treated as Ewald sums, with a constant negative background charge to compensate the positive total charge due to the positron.

There are also systematic finite-size errors in energy arising from the positron interacting with its periodic images. In metallic systems these errors should be small when the simulation cell is large compared to the Thomas-Fermi screening length. In semiconductors the error decreases with increasing simulation cell size as $v_M/2\epsilon$, where $v_M$ is the Madelung constant of the simulation cell that falls off with increasing number of atoms as $1/N^{1/3}$ for a given cell shape, and $\epsilon$ is the dielectric constant \cite{needs2020}.

Norm-conserving, nonlocal Dirac-Fock average relativistic effective pseudopotentials (AREP)\cite{trail2005,trail2005_2} are mainly used to approximate the ion cores. The positron-nucleus interactions are calculated with inverted electron pseudopotentials. We also computed lifetimes in Si with SJ wavefunctions using effective core potential (ECP) \cite{bennett2018} pseudopotentials with 2 electrons within the frozen core, and performed an all-electron SJ calculation for Li, enforcing cusp conditions on the electron and positron orbitals by adding short-ranged functions \cite{ma2005}.

The electron-positron annihilation results from the overlap of electrons and the positron in the many-body wave function \cite{gribakin2010}. The annihilation rate for 2$\gamma$ annihilation is (second form given in units of ns$^{-1}$)
\begin{align}
\label{equation: annihilation rate formula}
\begin{aligned}
\Gamma & = \pi r_0^2 c \sum_{i=1}^{N_e}\frac{\langle \Psi| \hat{O}_i^s\delta (\mathbf{r}_i-\mathbf{r}_+)|\Psi \rangle}{\langle\Psi|\Psi\rangle} = 100.9\ g(0)\frac{N_e^\uparrow}{V},
\end{aligned}
\end{align}
where $r_0$ is the classical electron radius, $c$ is the speed of light \textit{in vacuo}, $N_e$ ($N_e^{\uparrow}$) is the number of individual (spin-up) electrons, $V$ is the volume of the simulation cell and $\hat{O}^s_i$ is the spin-projection operator to the singlet state of the positron-electron pair. $g(0)$ is rotationally and translationally averaged contact pair correlation function (PCF).
In the case of metals (here Li), we make an asymptotic correction \cite{drummond2010} and multiply the PCF with $N_e/(N_e-1)$ to ensure that the effective electronic density is unchanged far from the positron in the simulation cell. 

The system-averaged PCF $g(|\mathbf{r}_e-\mathbf{r}_p|)$ is sampled with QMC by binning electron-positron distances. The leading-order errors in $\Psi_{\text{VMC}}$ are removed by extrapolating the final result as $g(r)=2g_{\text{DMC}}(r)-g_{\text{VMC}}(r)$ \cite{ceperley1979}. The estimate of $g(|\mathbf{r}_e-\mathbf{r}_p|)$ has poor statistics near the contact region $|\mathbf{r}_e - \mathbf{r}_p|\approx 0$. We estimate the $g(0)$ by fitting a $N$th-order polynomial $p(r)=a_0-r+a_2r^2+...+a_Nr^N$ to $\log(g(r))$ in the range $0<r<r_{cut}$, so that $g(0)=\exp(a_0)$. By setting $a_1=-1$ in the polynomial we assure that the fitted $\exp(p(r))$ satisfies the Kimball cusp conditions \cite{kimball1973}. Supplemental Material describes the details of the fitting procedure.

With pseudopotentials, the annihilation and screening interactions due to core electrons are not considered. We calculate reference DFT results and estimate the annihilation rate $\Gamma_c$ due to core electrons using a number of enhancement functionals: Drummond \textit{et al.} LDA (D-LDA)\cite{drummond2011}, Boro\'nski-Nieminen LDA (BN-LDA)\cite{boronski1986}, Kuriplach-GGA (KUR-GGA)\cite{kuriplach2014}, and GGAs by Barbiellini \textit{et al.} from Ref. \cite{barbiellini1995} (B95-GGA) and \cite{barbiellini2015} (B15-GGA). Hence the total annihilation rate is $\Gamma = \Gamma_{QMC} + \Gamma_c$.

For C and Si, we used  $2\times 2\times 2$ and $3\times 3 \times 3$ face-centered cubic (fcc) simulation cells, inluding $16$ and $54$ atoms ($64$ and $216$ electrons with AREP pseudopotentials). For Si, also a $4\times 4\times 4$ fcc cell with $128$ atoms ($512$ electrons) was investigated. Cubic Li  $3 \times 3\times 3$ or $5 \times 5 \times 5$ cells had $54$ or $250$ atoms, with one valence electron per atom. AlN was modeled with $2 \times 2\times1$ and $3\times 3\times 2$ hexagonal primitive cells, resulting in $16$- and $72$-atom ($64$- and $288$-electron) supercells. ECP pseudopotential and all-electron calculations had $3$ times more electrons per atom than in AREP simulations. 

We have also studied the convergence of the positron relaxation energy, defined as the energy difference between a system with and without the positron, $E_{r}=E_+-E_-$.

\begin{figure*}[!htb]\hspace*{-1.cm}
\includegraphics[scale=.45]{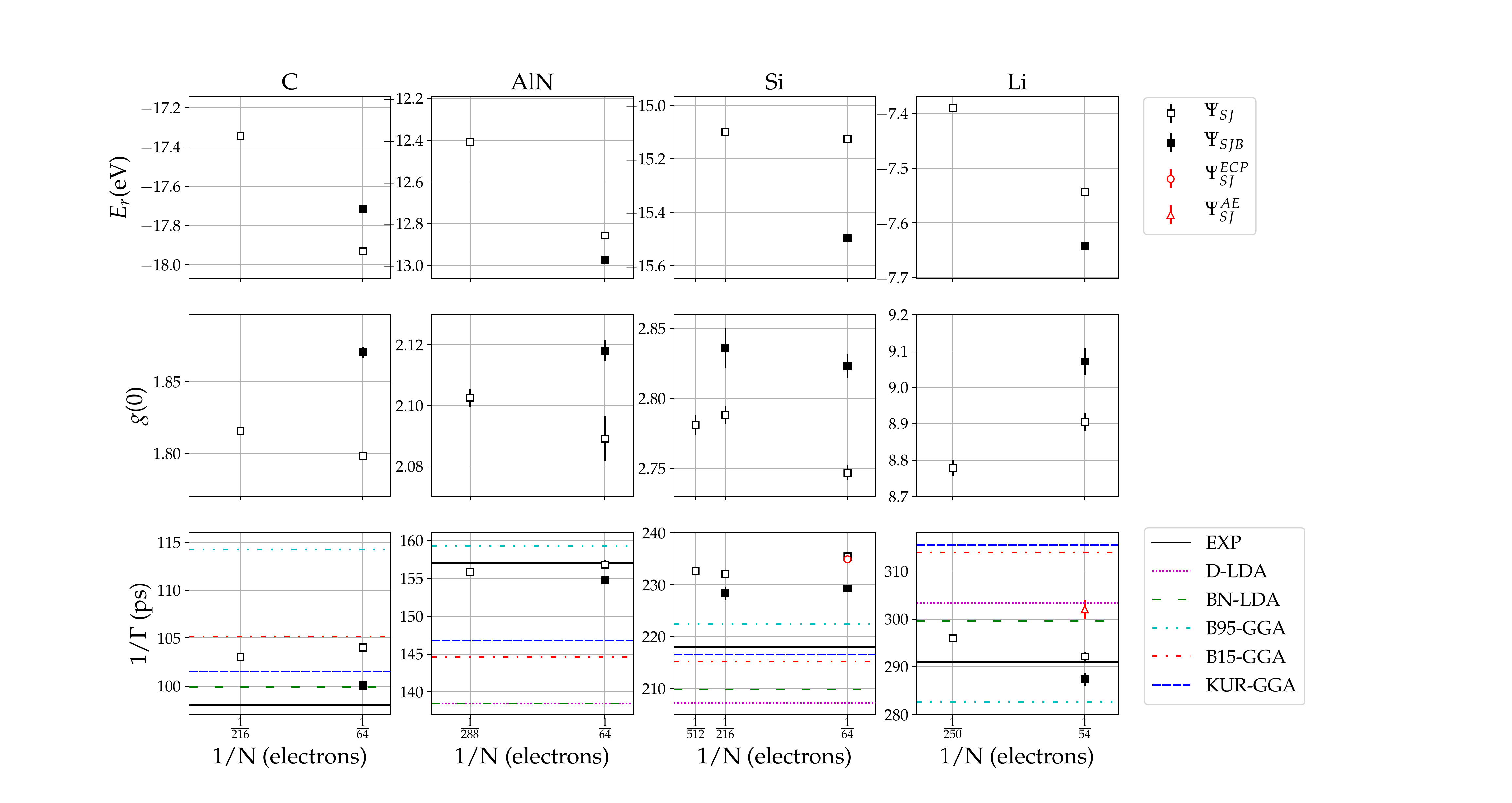}
\caption{\label{fig: results} Positron relaxation energies (top row), contact pair correlation functions (center row), and lifetimes (bottom row) for C, AlN, Si, and Li. The figure shows twisted SJ (empty square) and SJB (full square) wave function results with AREP pseudopotentials as well as the ECP pseudopotential (red circle) and all-electron SJ results (red triangle) as a function of the inverse of the number of electrons in the simulation. The ECP-pseudopotential and all-electron results are compared against AREP results with the same cell sizes. Monte Carlo errorbars are shown for each result. We present the lifetime estimates against experiment (black solid line)  for C \cite{pu2000}, AlN \cite{tuomisto2008}, Si \cite{makinen1992}, and Li \cite{Weisberg1967}. We also computed DFT lifetime estimates with different positron correlation functionals (dash-dotted lines): D-LDA (violet), BN-LDA (green), B95-GGA (cyan), B15-GGA (red), and KUR-GGA (blue). }
\end{figure*}

Figure \ref{fig: results} shows the QMC results of relaxation energies, $g(0)$ and lifetimes as a function of electron number. The Monte Carlo errors are shown in the errorbars. Lifetimes from DFT and experiment are also included. Twist averaging was done in a $\Gamma$-centered $4\times4\times4$ grid in the irreducible wedge of the Brillouin zone of the supercells. The backflow function was optimized separately for each twist, but with SJ wave functions the Jastrow factor optimized in the $\Gamma$-point was used for all of the twists. B15-GGA was used to approximate $\Gamma_c$. Other functionals gave mainly similiar results (see Supplemental Material). ECP pseudopotential and all-electron results are shown against AREP results with the same number of atoms.

In AlN, Si and Li the backflow decreases the relaxation energy by $100-400$ meV, and increases it by $200$ meV in C. The larger cell size increases the relaxation energy by approximately $600$ (C), $400$ (AlN), $25$ (Si) and $150$ (Li) meV. The backflow increases the $g(0)$ values. Only in Si we see convergence in PCF with respect to cell size, but changes in $g(0)$ and lifetimes between different cell sizes of the same material are small.

In the largest cells with SJB wave functions, QMC with AREP pseudopotentials overestimates the experimental lifetimes by $2$ and $10$ ps in C and Si and underestimates them by $2$ and $4$ ps in AlN and Li. The larger cells with SJ wave functions decreased the lifetimes in C, AlN  and Si by $1$ ps and increased by $4$ ps in Li.
In Si, SJ ECP and core-corrected AREP pseudopotential results agree within errorbars.
The all-electron lifetime of Li provides a $10$~ps increase to the core-corrected AREP result.

We studied the impact of atomic vibrations to lifetime values in Si. We used DFT to calculate the dynamical matrix in a cubic $64$-atom simulation cell and diagonalized it to obtain the eigenfrequencies and modes of the atomic vibrations. At temperatures of $0$ and $300$~K , we sampled a set of occupied states out of the Boltzmann distribution. Displacements were sampled with the Neumann algorithm based on the occupied states. By repeatedly occupying vibrational states and sampling atomic displacements we produced a set of $100$ atomic configurations distributed according to the occupied phonon modes. The average DFT lifetimes increased $2(1)$ ps compared to the lifetime of the static structure(see Supplement).  

The SJB results with AREP pseudopotentials and the reference experimental values are gathered in Table~\ref{table: best qmc lifetimes}. See Supplemental Material for QMC results with core corrections by different DFT functionals.

\begin{table}[ht]\caption{\label{table: best qmc lifetimes}The largest-cell SJB lifetime results obtained with AREP pseudopotentials with ($1/\Gamma$) and without ($1/\Gamma_{QMC}$) core corrections against experimental lifetimes. }
\begin{ruledtabular}
\begin{tabular}{lccr}
Lifetime (ps) & $1/\Gamma_{QMC}$ & $1/\Gamma$ & \textrm{Exp.\ reference}\\
\colrule
C   & 101.7(2) & 100.1(2)  &  98 \cite{pu2000}\\
AlN & 165.1(3) & 154.8(3)  & 157  \cite{tuomisto2008}\\
Si  & 237(1) & 228(1)    & 218 \cite{makinen1992}\\
Li  & 320(1) & 287(1)    & 291 \cite{Weisberg1967}\\
\end{tabular}
\end{ruledtabular}
\end{table}

The SJB lifetime values in C and AlN match almost perfectly with experimental values. Li matches also very well, but the all-electron SJB result could be overestimating the experimental value by $7$--$10$ ps, based on the SJ results, although it has to be noted that measurements on Li in the literature are scarce. The overestimation in Si (and Li) might result from the omission of relativistic and bound-state effects in Eq.~(\ref{equation: annihilation rate formula}) (both beyond the scope of the present Letter), finite size effects, fixed node errors, core electron approximations or vibrational effects.
Finite-size effects have been studied above. Because the backflow decreases the lifetime estimate, the QMC results are not converged with respect to the variational freedom in the wave function, and further improvements might be obtained by e.g. multideterminant wave functions. The ECP and all-electron calculations with SJB wave functions cannot be computed with current computing resources.

On average QMC shows better match with experimental lifetimes than DFT for this set of test materials. The mean-square error of QMC predictions against experimental values is $31.3(8)$ ps$^2$, as opposed to the best performing DFT functional B95-GGA with corresponding value of $89.4$ ps$^2$ (see Supplemental Material).  The QMC method is parameter free whereas all of the GGA functionals apart from B15-GGA involve one semiempiric parameter, determined by fitting to experimental data~\cite{barbiellini1995,kuriplach2014}. The LDA construction is unique and parameter-free but does not work for the positron lifetime. 

In conclusion, we have succesfully simulated electron-positron wave functions  and computed the positron lifetimes in crystalline C, Si, Li, and AlN with QMC. We studied finite-size effects, different pseudopotentials and included an all-electron calculation. The possibility of vibrational effects greatly affecting lifetimes of thermalized positrons in Si was ruled out.

The results prove that positron lifetime spectroscopy can benefit from the support of parameter-free many-body methods such as QMC calculations. The largest simulation cells in this study are applicable to vacancy calculations, and thus the presented method can be applied to support all fields of modern positron annihilation spectroscopy. 

\begin{acknowledgments}
We acknowledge the generous computational resources provided by CSC (Finnish IT Centre for Science) and the use of DECI resource Archer based in Edinburgh, UK, with support from the PRACE aisbl. This work was partially supported by the Academy of Finland grants No. 285809, 293932, 319178, 334706, and 334707.
\end{acknowledgments}

\end{document}


\title{Supplemental Material: Quantum Monte Carlo Study of Positron Lifetimes in Solids}

\author{K. A. Simula}
\author{J. E. Muff}
\author{I. Makkonen}\affiliation{Department of Physics, P. O. Box 43, FI-00014 University of Helsinki, Finland
}

\author{N. D. Drummond}
\affiliation{
Department of Physics, Lancaster University, Lancaster LA1 4YB, United Kingdom 
}

\date{\today}

\onecolumngrid

\maketitle

\onecolumngrid

\appendix

\onecolumngrid

\subsection{Preparation of trial wave functions}

We used the PWSCF package of Quantum ESPRESSO \cite{giannozzi2009} to solve the single-electron orbitals to be used in constructing the Slater determinants for the QMC wave functions. For each system, we studied the sufficient plane-wave cutoff with $1$ meV/atom convergence criterion. With this criterion, the all-electron simulation for Li required unpractically large plane-wave cutoffs. We then studied the convergence of the DMC energy, with cusps imposed on the orbitals by adding short-ranged functions \cite{ma2005}, for the Li cell as a function of the DFT plane-wave cutoff and chose the plane-wave cutoff to be used in further calculation based on the $1$ meV/atom convergence criterion. The chosen cutoff energies are reported in Table~\ref{table: plane-wave cutoffs}.

\begin{table*}[!h]

\caption{\label{table: plane-wave cutoffs} The used plane-wave cutoffs in the PWSCF calculations.}
\begin{tabularx}{\textwidth}{XXXXXXX}
\hline\hline
System&C AREP & AlN AREP & Si AREP & Si ECP & Li AREP & Li all-electron \\
\hline 
Cutoff (eV) & 2900 & 5400  & 1400 & 12000 & 820 & 19000 \\
\hline \hline
\end{tabularx}
\end{table*}

For C and Si we used a $2$-atom face-centered cubic, for AlN a $4$-atom hexagonal and for Li a $2$-atom cubic unit cell. To prepare trial wave functions for $n\times m \times l$ simulation cell, the orbitals were solved by applying a $n\times m \times l$ $\mathbf{k}$ vector grid in the unit cell simulation. The cubic simulation cell with $8$ Si atoms and AREP pseudopotentials had the same convergent cutoff as the fcc cell. 

After we had solved the one-electron orbitals with PWSCF, we solved the positron orbital within the potential of the obtained electron density by using our two-component DFT positron simulation package, and added it to the QMC trial wave function.The trial wave function given by PWSCF and atsup was represented as Slater determinant of the electron and positron orbitals, which were in turn represented by plane-wave sums before the blip conversion.

\subsection{Wave function optimizations}  

We used Jastrow  exponent, $J$, containing polynomial electron-electron and electron-positron terms $u$, electron-nucleus and positron-nucleus terms $\chi$, and electron-electron-nucleus or electron-positron-nucleus terms $f$. We also used backflow functions with corresponding polynomial particle-particle and particle-nucleus terms. Table \ref{tab:number of parameters in j and b} shows the number of parameters optimized for the Jastrow factor and the backflow function with different materials. The numbers in the table are the total number of parameters in the Jastrow factor, so that the parameters in different particle groupings are summed together. Optimizations with higher number of parameters were not found to decrease the system energy.

The optimizations of the Jastrow factors were performed with variance minimization, and the backflow was optimized with energy minimization. With ECP pseudopotentials and in all-electron simulations the variance minimization led to acceptance rates of $\approx 57 \%$ for the positron in the VMC simulations. The use of energy minimization fixed the acceptance rate to $50 \%$ and lowered the obtained energy and errorbars, and was hence chosen for the computations with ECP pseudopotentials or all-electron simulations. 

\begin{table}[!htpb]
\caption{\label{tab:number of parameters in j and b}The number of parameters in the Jastrow factor and backflow function. }
\begin{ruledtabular}
\begin{tabular}{lccr}
Jastrow & $u$ & $\chi$ & $f$ \\
\hline
C & 48 & 24 & 78 \\
Si & 60 & 24 & 78\\
AlN & 36 & 24 & 78 \\
Li & 24 & 16 & 24 \\
\hline
Backflow & $\eta$ & $\mu$ &\\
\hline
C& 18 & 12&\\
Si& 12 &8 & \\
AlN& 18& 12& \\
Li & 12 & 8 &
\end{tabular}
\end{ruledtabular}
\end{table}

\begin{figure}[!htb]\hspace*{-.5cm}
\includegraphics[scale=.59]{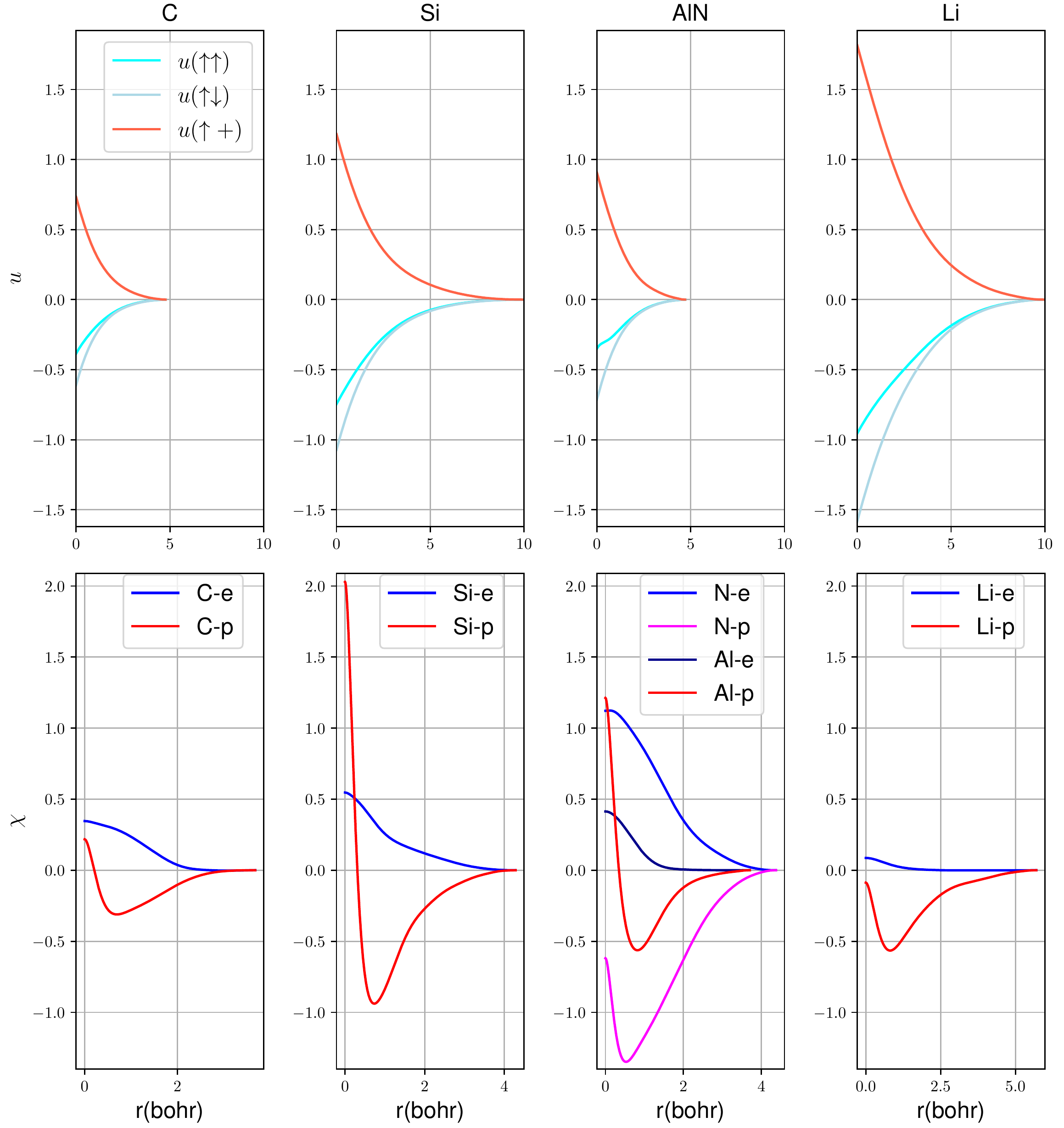}
\caption{\label{fig: opt. jastrow} The forms of the $u$- and $\chi$-terms in the Jastrow factor for Si, C, AlN, and Li.}
\end{figure}

Figure \ref{fig: opt. jastrow} shows the $u$- and $\chi$-terms for optimized SJB systems of different materials. The Jastrow factors are taken from 16-atom C,  54-atom Si, 16-atom AlN, and 54-atom Li systems, as these systems were found to give convergent positron relaxation energies. In all of the cases we have approximately $u_{\uparrow\downarrow}=-u_{\uparrow +}$. We can see that the strength of the correlation effects captured by the $u$-term increases as C-AlN-Si-Li. $\chi$-term shows that the electron-nucleus correlations are always described by a monotonically decreasing function of the particle-nucleus distance. The correlations between positron and the nuclei have attractive maximum values at contact in case of C, Si and Al atoms, but the N and Li atoms seems to have repulsive correlations with the positron at all distances. The correlations in the $\chi$-term for positron-nucleus correlation experience a minimum at distances $0.7$(C), $0.73$(Si), $0.8$(Al), $0.5$(N), and $0.81$(Li) atomic units. Clearly the strongest minimum is experienced in the case of N. One should note that the particle-nucleus correlations are very small for C  and Li when compared against the other atoms studied. Also the positron-Si correlations are very large at contact: ten times the value of contact correlations between positron and C and twice the value of the contact correlation of positron and Al, even though the electron-nucleus correlations are larger for Al than Si.        

\subsection{Vibrational effects on positron lifetime}

\begin{figure}[!htb]\hspace*{-1cm}
\includegraphics[scale=.43]{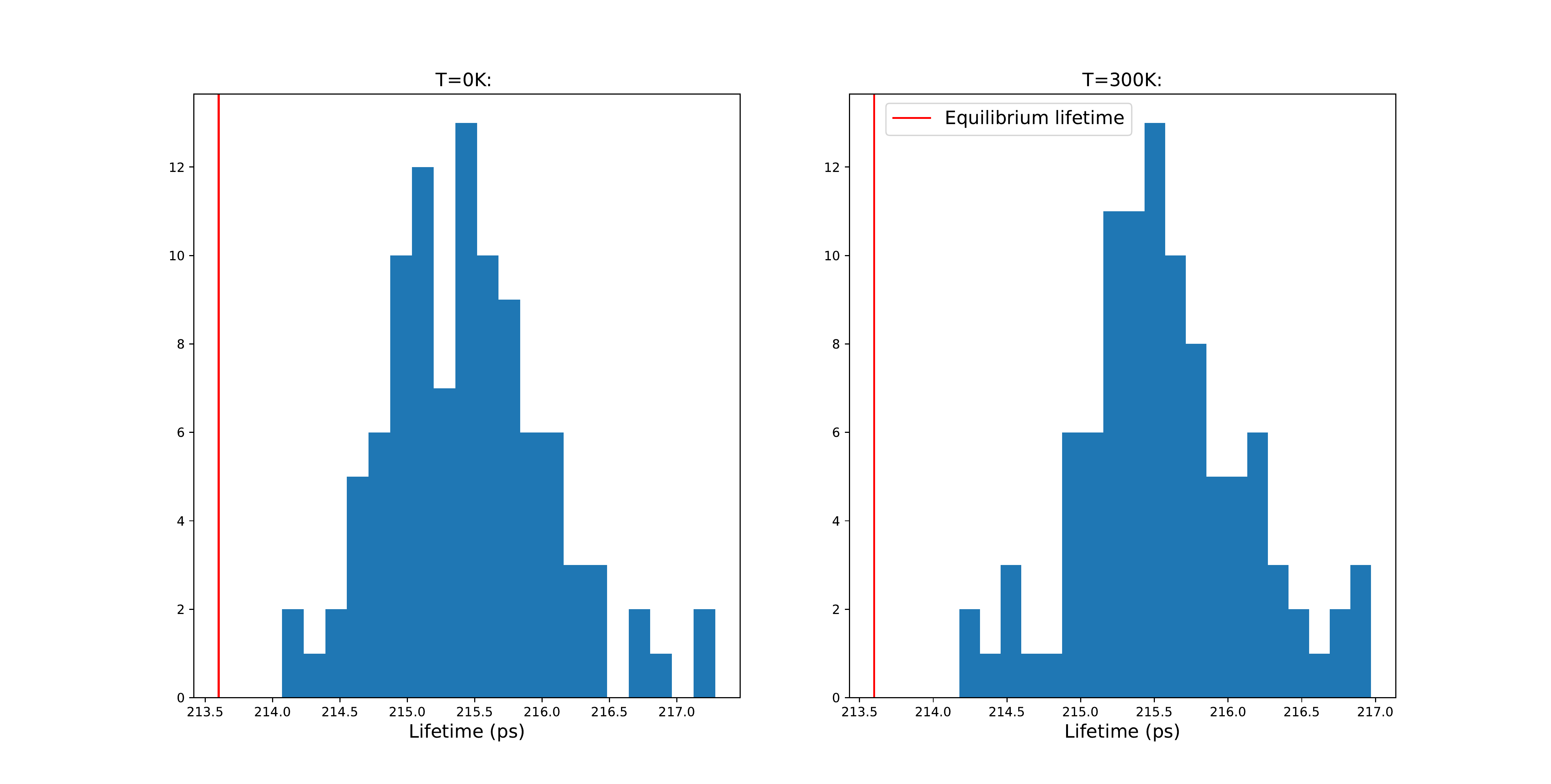}
\caption{\label{fig: si lifetimes} Distribution of BN-LDA positron lifetimes of bulk Si in diamond structure at $T=0$ (left) and $T=300$ K (right).}
\end{figure}

In order to solve the dynamical matrix and vibrational normal modes for a 64 atom cubic Si cell we used the PWSCF package \cite{giannozzi2009} with the PBE functional. First, we optimized the plane-wave cutoff within the desired accuracy, then we optimized lattice vectors to such as to give minimal energy with PBE. We used $\mathbf{k}$ grids with $3\times 3\times 3$ points.

After optimizing the numerical parameters of the density functional calculation, we made a phonon calculation. We solved the vibrational modes and frequencies with Quantum Espresso's Phonon-package \cite{giannozzi2009}. Then we produced $100$ PWSCF input files with different atomic displacements in $2$ different temperatures, $0$ and 300 K, and computed the lifetimes for all of the displacements. The BN-LDA was used to describe the electron-positron correlations..

Figure \ref{fig: si lifetimes} shows the lifetime distributions and the mean lifetimes and respective statistical errors. As can be seen, the vibrations have the effect of increasing the lifetime by $\sim$2 ps: while the LDA lifetime for an equilibrium atomic configuration is $213.6$ ps, the lifetimes with vibrational effects are $215.4$ ps ($0$ K) and $215.6$ ps ($300$ K), with standard deviations of $0.6$ ps at both temperatures. 

\subsection{Total energies}

\begin{table}[H]
\begin{ruledtabular}
\begin{tabular}{lcccccr}
 & $E^\text{VMC}_\text{SJ}$ (eV) & $E^\text{VMC}_\text{SJB}$ (eV) & $E^\text{DMC}_\text{SJ}$ (eV)& $E^\text{DMC}_\text{SJB}$ (eV) & Variance(SJ) (eV) & Variance (SJB) (eV) \\
\hline
C & $-2464.544(8)$ & $-2469.032(8)$ & $-2470.56(2)$ & $-2472.12(2)$ & $41.14(8)$ & $40.8(3)$ \\
AlN  & $-2647.03(1)$ & $-2650.84(1)$ & $-2655.12(2)$ & $-2655.93(2)$ & $42.2(1)$ & $52.0(3)$\\
Si & $-1705.536(4)$ & $-1709.362(4)$ & $-1709.44(2)$ & $-1710.90(1)$ & $14.45(5)$ & $11.08(5)$ \\
Li & $-387.820(3)$ & $-390.56(1)$ & $-390.76(1)$ & $-392.070(8)$ & $4.41(2)$ & $4.2(1)$
\end{tabular}
\end{ruledtabular}
\caption{\label{table: total energies} Total VMC and DMC energies per simulation cell of the SJ and SJB wave functions in C, AlN, Si, and Li, obtained from a calculation in a $64$-electron ( $54$ in Li) simulation cells with a positron. The two columns in the right show the variances of local energy in the VMC calculations. The statistical error is reported for the final decimal in the brackets.  }
\end{table}

Table \ref{table: total energies} shows the total energies of the $\Gamma$-twist in the $64$-electrons ($54$-electron in Li) supercells of the simulated systems including a positron, computed with Slater-Jastrow (SJ) and Slater-Jastrow-backflow (SJB) wave functions using VMC and DMC. For VMC also the variances of the local energy are reported. The values in Table \ref{table: total energies} have been obtained with the AREP pseudopotentials.

\subsection{Twist averaging of energies}

The twist-averaged energies are obtained by fitting \cite{needs2020}
\begin{equation}
E_{\text{QMC}}(\mathbf{k}_s)\approx E_{\text{QMC}}(TA)+b(E_{\text{DFT}}(\mathbf{k}_s)-E_{\text{DFT}}(\text{fine}))
\end{equation}
to the QMC energy data $E_{\text{QMC}}(\mathbf{k}_s)$ as a function of twist $\mathbf{k}_s$ for a given size of superell. $E_{\text{DFT}}(\mathbf{k}_s)$ is the DFT energy with the $\mathbf{k}$ point grid corresponding to the QMC calculation at twist $\mathbf{k}_s$. $E_{\text{DFT}}(\text{fine})$ is a DFT calculation with a fine $\mathbf{k}$ grid. The twist-averaged energy $E_{\text{QMC}}(TA)$ and $b$ are fitting parameters.

\subsection{Polynomial fits to QMC data}

\begin{figure}

\includegraphics[scale=.5430]{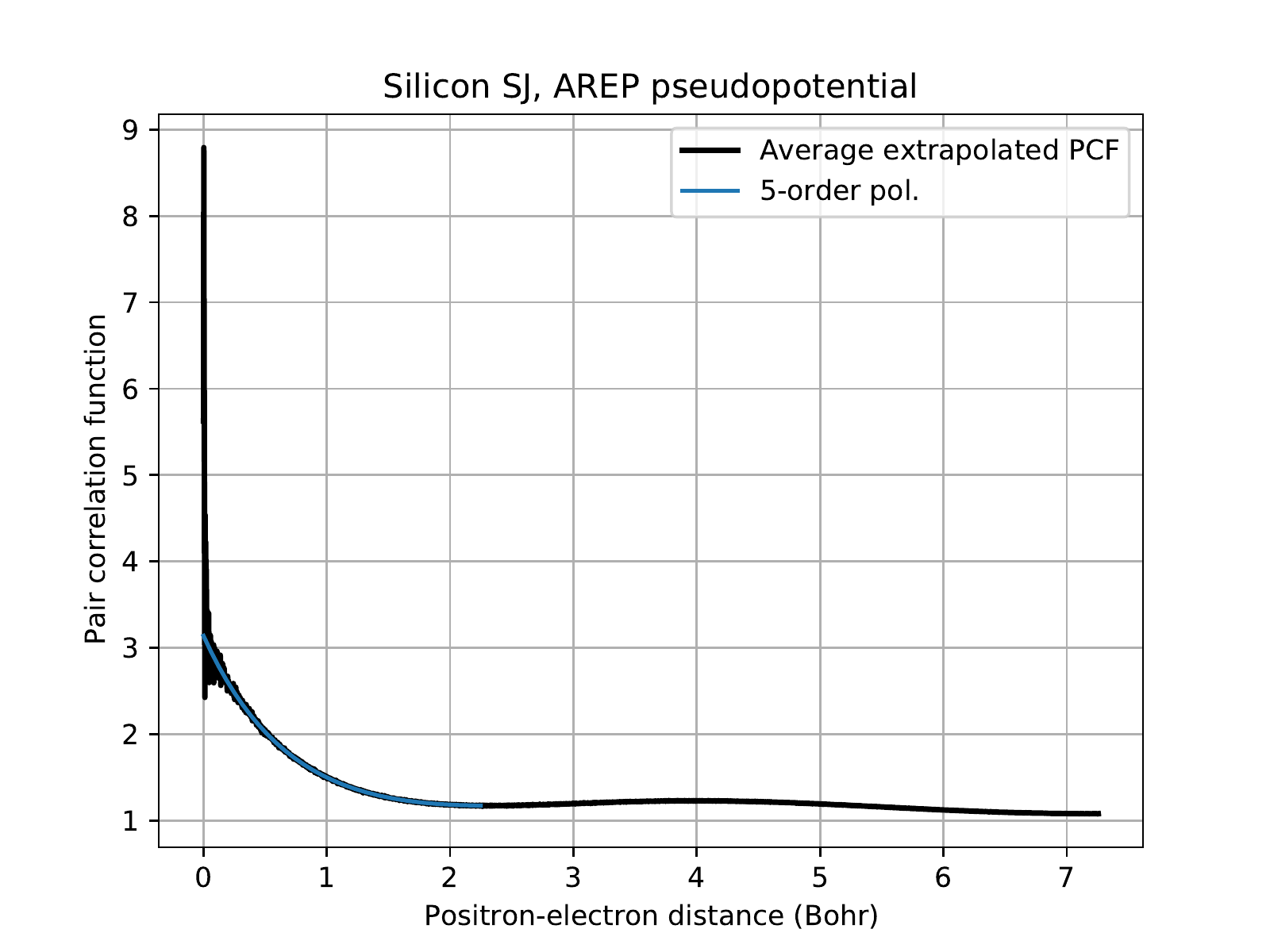}
\includegraphics[scale=.542]{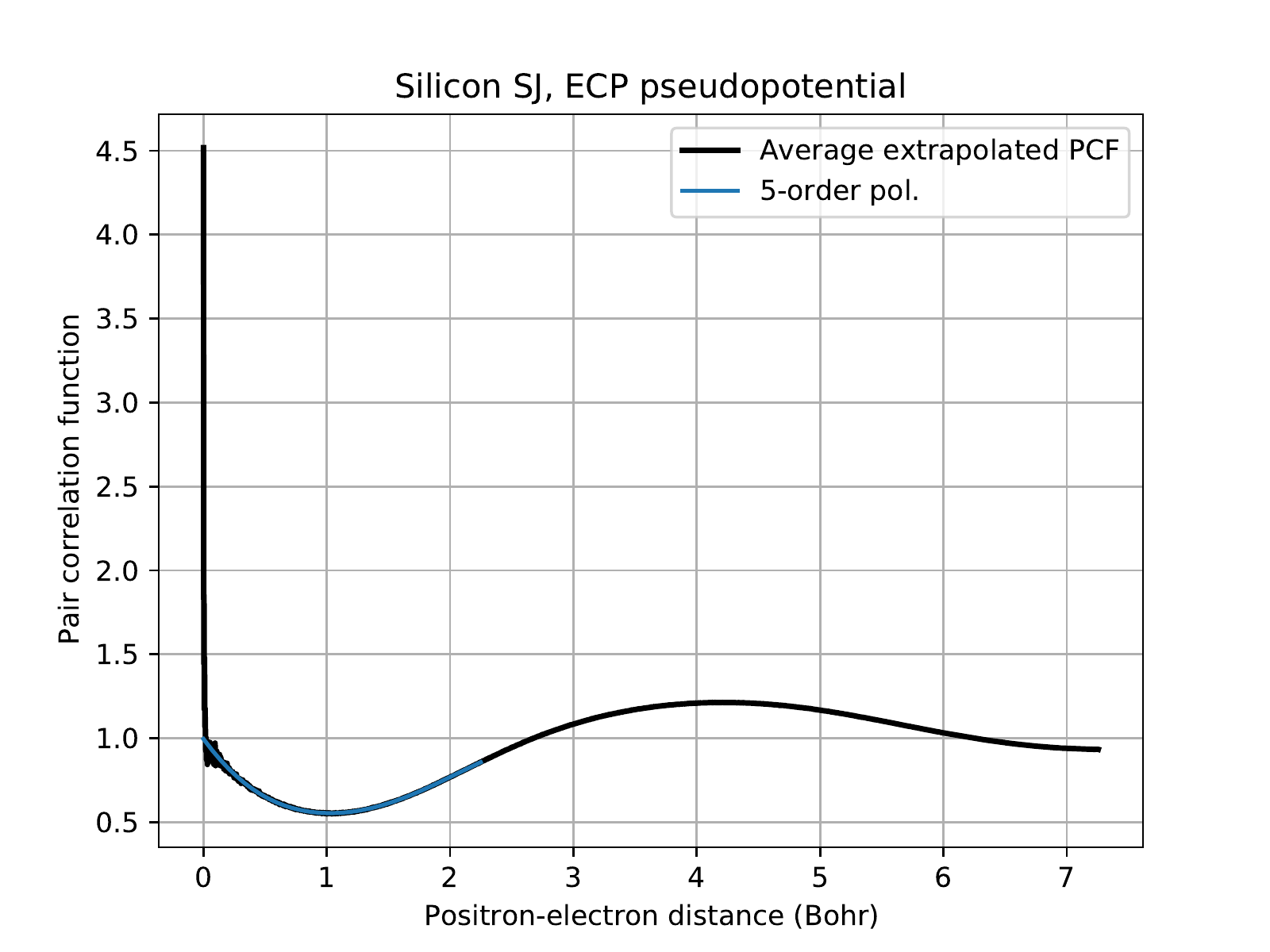}
\caption{\label{figure: pcf fits}Average PCF histograms, obtained by extrapolation of VMC and DMC PCF data at each twist, against the average of exponentiated fits with $5$th-order polynomials. The PCF data presented here are from QMC simulations with Si SJ wavefunctions in a $16$-atom fcc simulation cell, with both AREP (left) and ECP (right) pseudopotentials.}
\end{figure}

The PCFs were accumulated $2$ times for each of the twists to get independent data. Thus for $N_t$ twists we had $2N_t$ independent PCF histograms $g^i(r)$ from the VMC-DMC extrapolation. The PCF data was noisy and contained divergences close to zero positron-electron distance region, and before extrapolation we removed negative values from the extrapolated histograms. To obtain the lifetime estimates, polynomials $p^i(r)=a^i_0+a^i_1x+...+a^i_Nx^N$ were fitted with a chosen fitting range to each of the individual and independent $\log(g^i(r))$ histograms by using the non-linear least-squares based Levenberg-Marquardt algorithm \cite{marquardt1963,levenberg1944}. The final estimate of the PCF at zero was taken to be the average of the exponentiated values $\frac{1}{2N_t}\sum_{i=1}^{2N_t}\exp(a^i_0)$ at zero distance. This way we can both make the results more robust against noise near zero distances and obtain an error estimate for the contact PCFs and lifetimes. Figure \ref{figure: pcf fits} shows the PCF averaged over all accumulated data $\frac{1}{2N_t}\sum_{i=1}^{2N_t}p^i(r)$ against the averaged and exponentiated fits $\frac{1}{2N_t}\sum_{i=1}^{2N_t}\exp(p^i(r))$, as well as the average fits with $5$th-order polynomials fitted in range $0$-$2.25$ Bohr, for the SJ Si results with AREP (left) and ECP (right) pseudopotentials.

\begin{figure}
\includegraphics[scale=.71]{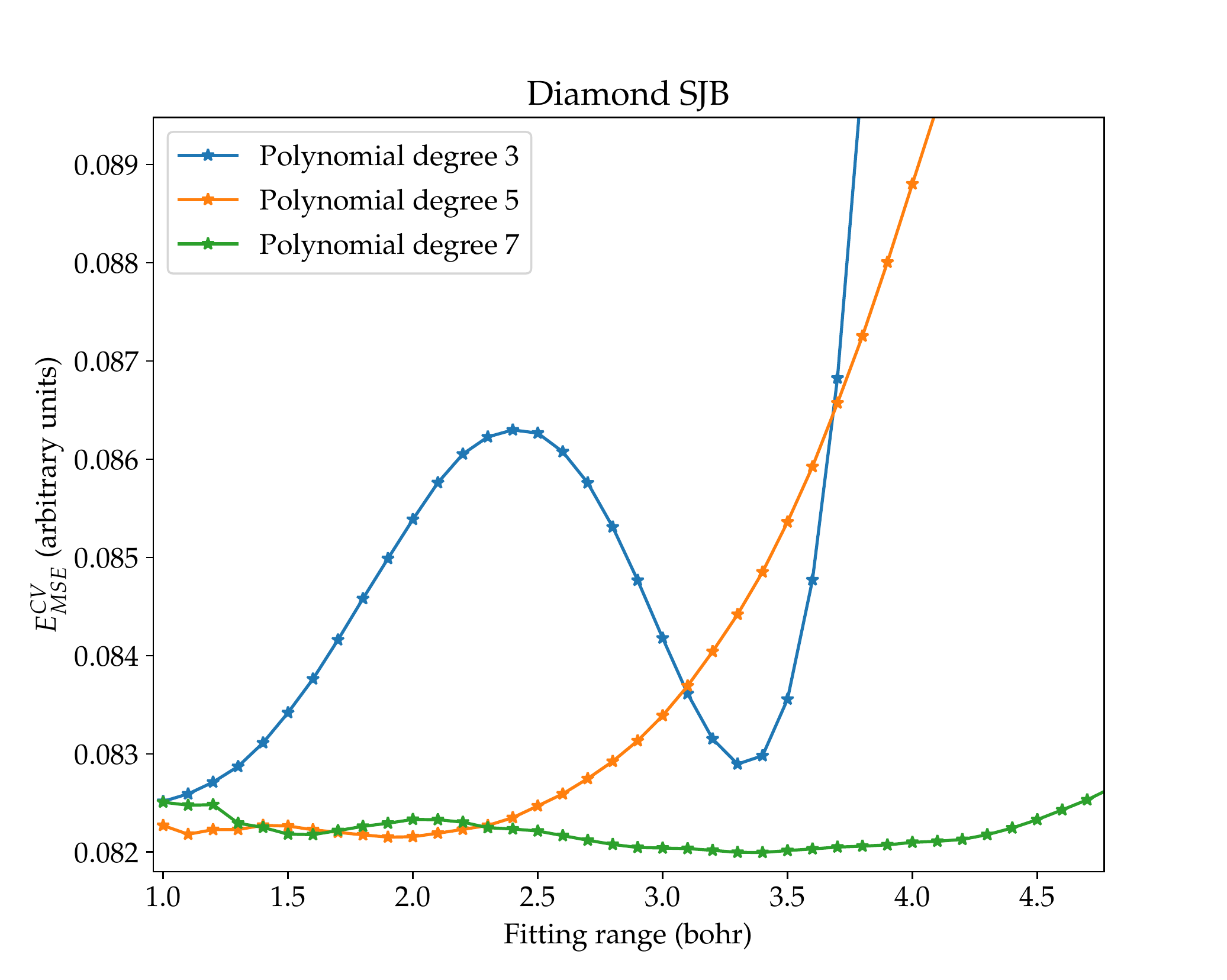}
\caption{\label{figure: cross validation silicon SJB example} Results of a cross-validation analysis on the fitting parameters with different polynomial orders and fitting ranges for a C $16$-atom simulation cell and SJB wave function.  }
\end{figure}

Before fitting we had to choose the polynomial order and fitting range. For this purpose we performed a cross-validation analysis on the fitting parameters by comparing mean-squared error (MSE) estimates between individual fits against PCF data not used in the fitting. With a candidate polynomial order and fitting range, we fitted a polynomial $p^i(r)$ to each $2N_t$ PCF data $g^i(r)$. Then for each fit we computed a MSE estimate as $E_{MSE}^i=\frac{1}{N_x}\sum_{n=1}^{N_x}\left[ g^i(x_n) - p^j(x_n) \right]^2$, where $N_x$ is the number of columns in the PCF histograms, and $x_n$ is the distance related to $n$th histogram. The index $j$ denotes the PCF histogram corresponding to the same twist as $g^i$, but obtained from an independent QMC simulation. The final cross-validation MSE estimate was obtained as $E^{CV}_{MSE}=\frac{1}{N_t}\sum_{i=1}^{N_t}E_{MSE}^i$. We computed this estimate with varying polynomial order and fitting range for each material, cell size, pseudopotential (including the all-electron calculation), and wave function approximation. An example of our cross-validation results on silicon with SJB wave function in the $54$-atom simulation cell can be seen in Fig. \ref{figure: cross validation silicon SJB example}.

The lowest  $E^{CV}_{MSE}$ were obtained in C, AlN and Si with $7$th-order polynomials at fitting ranges of $3.5$-$5$ Bohr. However, with $5$th-order polynomials and fitting ranges of $2$ (C), $2.2$ (AlN) and $2.5$ (Si), the cross-validation errors were within $1$\% of the lowest value with $7$th-order polynomial. In Li, the lowest cross-validation error values were obtained with $5$th-order polynomials at all distances, with the fitting range of $\approx 5$ Bohr giving the lowest value, with slight variation between wave function approximations, simulation cell sizes and core electron approximations. Again, the $5$th-order polynomial with fitting range of $2.25$ Bohr gave cross-validation errors that were within $1$\% of the lowest value. 

We wanted to use the same method for all of our PCF data for consistency. Based on the findings listed above we chose to use $5$th-order polynomial and fitting range of $2.25$ Bohr, as this provided robust results with all materials, gave a cross validation error within $1$\% of the best value in all systems, and was a fit to relatively small distance and polynomial order hence preventing any catastrophic errors from occurring. Results obtained by fitting in each system with polynomial order and fitting range giving the minimal $E^{CV}_{MSE}$ provided to be the same with our chosen method, with respect to Monte Carlo errorbars.

\subsection{Core corrections}

Table  \ref{table: valence and full lifetimes} presents lifetime estimates (valence annihilation only) of DFT and QMC calculations, as well as the core-corrected AREP QMC estimates with different density functionals. The experimental lifetimes are shown for comparison. The C and AlN estimates are taken from the $64$-electron, the Si estimate from the $216$-electron and the Li estimate from the $54$-electron simulation cell with twisted SJB wave functions. Li is an exception, as different density functionals both under- and overestimate the QMC result. For all of the materials except Li, the core-corrected QMC lifetime estimates seem to be relatively robust against the choice of functional for the core correction, as the lifetime estimates are with different core correction functionals within the range of $\sim 2$~ps, which is only a little larger than the statistical error on the results.

\begin{table*}[!h]

\caption{\label{table: valence and full lifetimes}Lifetimes computed with DFT and QMC with AREP pseudopotentials against experimental values. The predictions with only valence electrons are considered for different DFT functionals and QMC for C ($ \tau_{\text{valence}}^{\text{C}}$), AlN ($ \tau_{\text{valence}}^{\text{AlN}}$), Si ($ \tau_{\text{valence}}^{\text{Si}}$) and Li ($ \tau_{\text{valence}}^{\text{Li}}$). Below the lines with valence-only lifetime estimates for each material the core-corrected lifetime predictions are presented, computed as $\tau_{full}=(\Gamma_c+\Gamma_v)^{-1}$, where the core annihilation rate is estimated with the DFT functional of the corresponding column, and the valence rate is from the QMC simulation.}
\begin{tabularx}{\textwidth}{XXXXXXXXXXXXXX}
\hline\hline
                                        & D-LDA    & BN-LDA   & B95-GGA  & B15-GGA  & KUR-GGA  & QMC       & Experiment\\
\hline
$ \tau_{\text{valence}}^\text{C}$(ps)   & 103.3    & 101.8    & 115.9    & 107.0    & 103.3    & 101.7(2)  & \\
$ \tau_{\text{full}}^\text{C}$(ps)      & 100.0(2) & 99.8(2)  & 100.4(2) & 100.1(2) & 100.0(2) &           & 98\cite{pu2000}\\
\hline
$ \tau_{\text{valence}}^\text{AlN}$(ps) & 147.7    & 148.1    & 166.9    & 153.6    & 155.5    & 165.1(3)  &\\
$ \tau_{\text{full}}^\text{AlN}$(ps)    & 153.7(3) & 153.3(3) & 157.7(3) & 154.8(3) & 155.4(3) &           & 157\cite{tuomisto2008}\\
\hline
$ \tau_{\text{valence}}^\text{Si}$(ps)  & 214.5   & 217.5     & 228.1    & 223.7    & 223.4    & 237(1)    &\\
$ \tau_{\text{full}}^\text{Si}$(ps)     & 228(1)  & 228(1)    & 231(1)   & 228(1)   & 229(1)   &            & 218\cite{makinen1992}\\
\hline
$ \tau_{\text{valence}}^\text{Li}$(ps)  & 351     & 347       &  301     & 353    & 353        & 320(1)    &  \\
$ \tau_{\text{full}}^\text{Li}$(ps)     & 280(1)  & 279(1)    & 299(1)   & 287(1) & 289(1)     &         & 291\cite{Weisberg1967} \\
\hline \hline
\end{tabularx}
\end{table*}

\section*{Theory vs. experiment errors}

\begin{table*}[!h]

\caption{\label{table: ME and MSE} Mean errors (ME) and mean square errors (MSE) of the theoretical lifetime predictions using different DFT functionals and QMC against experimental values. }
\begin{tabularx}{\textwidth}{XXXXXXXXXXXXX}
\hline\hline
                                & D-LDA    & BN-LDA   & B95-GGA  & B15-GGA  & KUR-GGA  & QMC     \\
\hline
ME (ps)                         &   11.3   &  9.3     &  7.8     & 11.1     & 9.9      & 4.6(4)  \\
MSE (ps$^2$)                    &   155.7  & 121.6    &   89.4   & 183.2    & 179.9    &  31.3(8)   \\

\hline \hline
\end{tabularx}
\end{table*}